  \documentclass{ifacconf}

  \usepackage{graphicx}      
  \usepackage{natbib}        

  \usepackage{bm}
  \usepackage{amsmath}
  \usepackage{mathrsfs}
  \usepackage{amssymb,amsfonts}
  \usepackage{mathtools}
  \usepackage{upgreek}
  \usepackage{hhline}
  \usepackage{float}
  \usepackage{siunitx}

\begin{document}
  \begin{frontmatter}

  \title{Acoustic-based Guidance for Automatic Docking of Holonomic AUVs $^\star$}
  
  

  \thanks[footnoteinfo]{This work was supported by the LARSyS FCT  funding (UID/50009/2025: DOI 10.54499/UID/50009/2025; LA/ P/0083/2020: DOI 10.54499/LA/P/0083/2020).}

  \author{Ravi Regalo\normalfont{,}}
  \author{David Cabecinhas\normalfont{, and}}
  \author{António Pascoal}
  
  \address{Institute for Systems and Robotics (ISR), Instituto Superior Técnico (IST), University of Lisbon, Portugal \\
  \{ravi.regalo, david.cabecinhas, antonio.pascoal\}@tecnico.ulisboa.pt}
  
  
  \begin{abstract}                

	This paper describes a system to automatically dock an AUV onto a docking station without precise knowledge of the position and orientation of the latter, in the presence of unknown ocean currents, using a fully acoustic sensing architecture.
	The system relies on a pair of Ultrashort Baseline sensors, one onboard the vehicle and one installed on a seabed-resident docking station, enabling operation in low-visibility environments where cameras are ineffective.
	Relative orientation is estimated by a nonlinear complementary filter on $SO(3)$, while an Extended Kalman Filter provides relative position, supplying pose estimates to a geometric controller on $SE(3)$ that executes the docking manoeuvre.
	The complete system is implemented in a dedicated software suite and validated in simulation and water trials.

\end{abstract}

  
\end{frontmatter}
\makeatletter
\def\ps@copyright{%
  \let\@mkboth\@gobbletwo
  \def\@oddhead{\@logohead}%
  \let\@evenhead\@oddhead
  \def\@oddfoot{%
    \raisebox{-8pt}[0pt][0pt]{%
      \parbox{\textwidth}{%
        \centering\normalfont\fontsize{10}{10}\selectfont
        \textcopyright{} 2026 the authors. This work has been accepted to IFAC for publication under a Creative Commons Licence CC-BY-NC-ND.}}}%
  \let\@evenfoot\@oddfoot
}
\makeatother

  \section{Introduction}\label{section:intro}

Autonomous exploration of the seabed is increasingly facilitated by affordable Autonomous Underwater Vehicles (AUVs), yet their endurance remains limited by battery capacity, constraining long-term and large-scale operations. Bottom-resident Docking Stations (DSs) equipped with power and data exchange interfaces are a promising solution to this problem by enabling periodic underwater battery recharging and data offloading to a base station, provided the AUV can robustly localize, approach, and dock onto the station. This brings several advantages such as reducing logistic costs and maximizing profits, increasing safety via continuous inspection of critical infrastructure, and advancing science through persistent biodiversity and environmental monitoring.

The majority of published work on underwater docking considers non-holonomic, torpedo-shaped AUVs approaching funnel-shaped DSs. Surveys such as~\cite{yazdani2020survey,esteba2021survey,liu2024review}, consistently report that the bulk of experimentally validated systems follow this pattern.
In these systems, terminal guidance is predominantly vision-based, see for example~\cite{park2009experiments} that employs visual servoing to light beacons or simple image features. Works such as~\cite{li2015auv} and~\cite{ghosh2016reliable} estimate the relative pose between the AUV and the DS from LED patterns and docking-plate markers in the last tens of metres. 
However, water turbidity and lighting constraints severely hinder the performance of purely optical solutions in realistic conditions. To mitigate these effects,~\cite{lin2022underwater} introduces electromagnetic guidance laws based on magnetic-dipole models for close-range homing, but these require dedicated hardware that is only useful for this purpose. 
Other approaches such as~\cite{palomeras2018auv} and~\cite{fan2019auv} rely on Ultrashort Baseline (USBL) sensors for positioning, but typically use acoustics only for mid-range navigation and then switch back to camera-based visual guidance for the terminal phase.

Considerably less work explicitly targets docking for fully actuated AUVs, despite their increased manoeuvrability and ability to decouple translation and heading. Available examples such as~\cite{maki2013docking},~\cite{palomeras2016auv} and~\cite{wang2021visual} still rely on camera-based terminal guidance. In these cases, cameras remain the primary sensor in the last approach, and acoustic information is either absent or employed during the initial approach phase only.
To the best of the authors' knowledge, no previous work has demonstrated a docking solution that relies solely on acoustic sensors for both mid-range and terminal phases while explicitly exploiting the control advantages of a fully actuated AUV, which is the gap addressed in this paper.

In the above context, an integrated solution for autonomous docking is derived, based on fully relative navigation using two USBL sensors, one installed on the AUV and one on the DS. The proposed solution does not require precise prior knowledge of the position and orientation of the DS and compensates for the influence of unknown ocean currents.
An acoustic communication link is established between the AUV and the DS, allowing the vehicle to access the measurements made by the USBL installed on the DS. The dual-USBL geometry is exploited to estimate the AUV's position and orientation with respect to the DS, enabling guidance and terminal docking even in low-visibility conditions where optical systems fail. 
The proposed methods are implemented in the Robot Operating System (ROS) and evaluated in a realistic simulation environment. They are then demonstrated experimentally using the Medusa AUV and a custom DS developed by the Dynamical Systems and Ocean Robotics (DSOR) group at Institute for Systems and Robotics (ISR), Instituto Superior Técnico (IST). Across these experiments, the proposed algorithms consistently enabled multiple consecutive successful docking manoeuvres under challenging environmental conditions, demonstrating the practical effectiveness of the approach for persistent subsea operations.

The remainder of this paper is organised as follows. Section~\ref{section:model} introduces the notation and vehicle model. Section~\ref{section:docking_strategy} formulates the docking problem and outlines the overall strategy adopted for its solution. Section~\ref{section:filter} details the dual-USBL relative navigation system and Section~\ref{section:control} describes the trajectory-tracking controller used in the terminal phase of the manoeuvre. Section~\ref{section:results} reports simulation and experimental results. Finally, Section~\ref{section:conclusion} concludes the paper and discusses directions for future work.
  \section{System Overview and Vehicle Modelling}\label{section:model}
%

\begin{figure}[b]
	\centering
	\includegraphics[width = 0.9\linewidth]{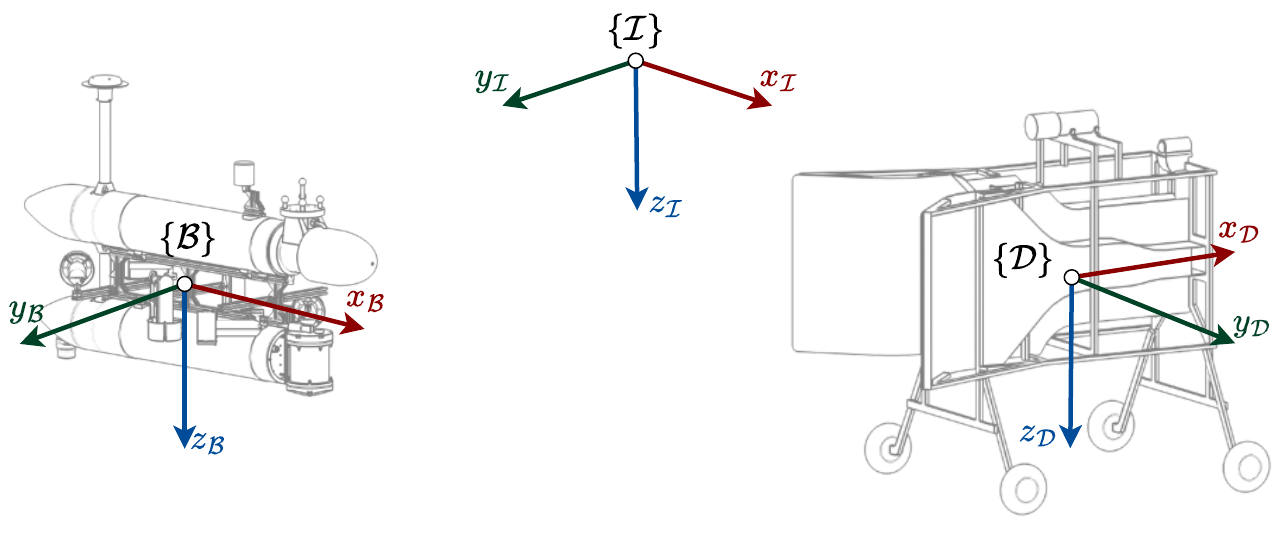}
	\caption{Adopted reference frames: inertial frame $\mathcal{I}$, vehicle body frame $\mathcal{B}$ and docking station frame $\mathcal{D}$.}\label{fig:ch3:reference_frame_notation}
\end{figure}
The adopted reference frames and notation follow~\cite{fossen2011handbook}. Represented in Figure~\ref{fig:ch3:reference_frame_notation}, the inertial frame $\mathcal{I}$ follows the North-East-Down (NED) convention, the body frame $\mathcal{B}$ has its $x$-axis aligned with the vehicle bow, and the DS frame $\mathcal{D}$ coincides with $\mathcal{B}$ when the vehicle is inside the DS.
The following notation will be used:
\begin{itemize}
    \item $\bm{\eta} = {[x \; y \; z \; \phi \; \theta \; \psi ]}^\top$: position  and orientation of $\mathcal{B}$ expressed in $\mathcal{I}$;
    \item $\bm{\nu}= {[u \; v \; w \; p \; q \; r]}^\top$: linear  and angular velocities of $\mathcal{B}$ with respect to $\mathcal{I}$, expressed in $\mathcal{B}$;
    \item $\bm{\tau} = {[\tau_u \; \tau_v \; \tau_w \; \tau_p \; \tau_q \; \tau_r]}^\top$: forces and moments applied to the vehicle, expressed in $\mathcal{B}$;
\end{itemize}

The adopted kinematic model is 
\begin{equation}
    \underbrace{
    \begin{bmatrix}
        \bm{\Dot{\eta}_1} \\
        \bm{\Dot{\eta}_2}
    \end{bmatrix}
    }_{\bm{\bm{\Dot{\eta}}}}
    =
    \underbrace{
    \begin{bmatrix}
         ^\mathcal{I}_\mathcal{B}R(\bm{\eta_2}) & \bm{0}_{3 \times 3} \\
        \bm{0}_{3 \times 3} & T(\bm{\eta_2})
    \end{bmatrix}
    }_{J(\bm{\eta})}
    \underbrace{
    \begin{bmatrix}
        \bm{\nu_1} \\
        \bm{\nu_2} 
    \end{bmatrix}
    }_{\bm{\nu}}
    ,
    \label{eq:ch3:kinematics}
\end{equation}
where $ ^\mathcal{I}_\mathcal{B}R(\bm{\eta_2})$ is orthogonal and characterizes the rotation from $\mathcal{B}$ to $\mathcal{I}$, and  $T(\bm{\eta}_2)$ is the Euler attitude transformation matrix. 

The dynamics of the AUV are described in compact form by
\begin{equation}
    M \bm{\dot{\nu}} + \bigl( C(\bm{\nu}) + D(\bm{\nu}) \bigr)\bm{\nu} + g(\bm{\eta}) = \bm{\tau} + \bm{\tau}_d,
    \label{eq:ch3:dynamics_general}
\end{equation}
where $M$ is the rigid-body and hydrodynamic mass matrix, $C_{RB}$ the rigid-body Coriolis and centripetal matrix, $D$ the linear and quadratic damping matrix, $\bm{g}$ the vector of restoring forces and moments, and $\bm{\tau}_d$ consists of external disturbances.

\begin{figure}[t]
	\centering
	\includegraphics[width = 0.8\linewidth]{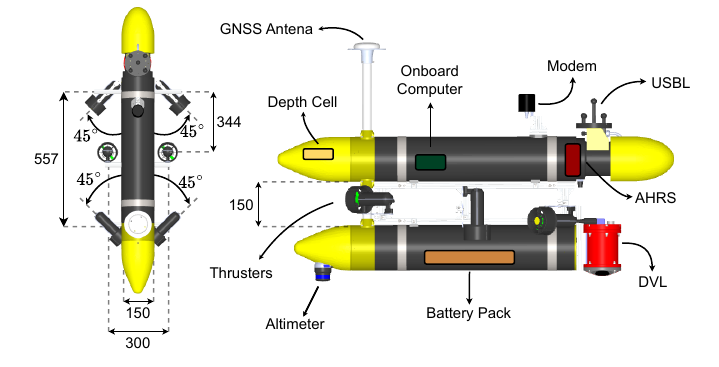}
	\caption{Hardware diagram of the Medusa vehicle, measurements are in \si{\milli\metre}.}\label{fig:model:medusa_hardware_diagram}
\end{figure}
\begin{figure}[t]
	\centering
	\includegraphics[width = 0.8\linewidth]{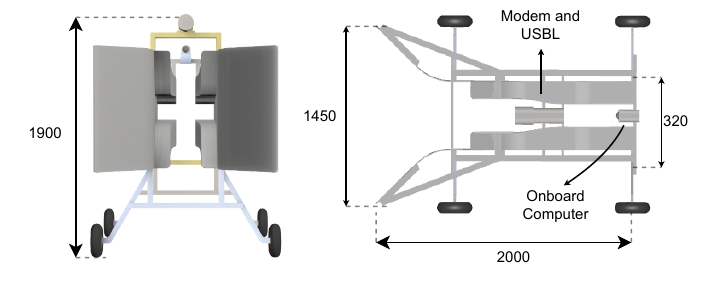}
	\caption{Hardware diagram of the docking station for the Medusa vehicle, measurements are in \si{\milli\metre}.}\label{fig:model:dock_hardware_diagram} 
\end{figure}
The Medusa AUV
is a fully-actuated platform in the horizontal plane, equipped with a sensor suite comprising GNSS, DVL, a depth sensor, an Attitude and Heading Reference System (AHRS), and an acoustic modem with an USBL (Figure~\ref{fig:model:medusa_hardware_diagram}).
The funnel-shaped DS, geometrically matched to the vehicle is also equiped with a modem and an USBL (Figure~\ref{fig:model:dock_hardware_diagram}). 
The USBL mounted on the DS is aligned with the opening, while the vehicle USBL and acoustic modem point upward. Due to the directional beam patterns of these sensors, blind zones arise above and behind the DS, restricting the region in which acoustic communication can be established.

Regarding communication protocol, an \textit{interrogation-reply} scheme is used: the AUV initiates the exchange and waits for a response from the DS. The range is measured based on the time delay between the transmission and reception of signals.

The two USBL systems provide reciprocal relative position measurements: the AUV measures the direction to the DS, while the DS measures the direction to the AUV. 
For simplicity, both measurements are expressed directly in the corresponding body frames, and the DS-side measurement is made available to the vehicle through acoustic communication. Each measurement is parameterized by the bearing $\beta$ and elevation $\alpha$. To avoid singularities, these angles are represented by the corresponding unit direction vector,
\begin{equation}\label{rbe_to_unit_los}
    \bm{u}(\beta,\alpha)=
    \begin{bmatrix}
        \cos\alpha\cos\beta \,\,\,\,& \cos\alpha\sin\beta  \,\,\,\,& \sin\alpha
    \end{bmatrix}.
\end{equation}
In what follows, let $\mathbb{P}_{\bm{u}}$ denote the projection onto the plane orthogonal to a unit vector $\bm{u}$, given by
\begin{equation}
    \mathbb{P}_{\bm{u}}\,\bm{x} = (I - \bm{u}\bm{u}^\top)\bm{x},
\end{equation}
which removes the component of $\bm{x}$ along $\bm{u}$.

  \section{Docking Strategy}\label{section:docking_strategy}

The docking problem considered in this work consists of guiding an AUV into a DS from an arbitrary initial condition where only an approximate inertial position of the DS is known and its orientation is unknown. The navigation and control framework used in the last step of this manoeuvre requires the vehicle to be within a region where acoustic communication with the DS is feasible, whereas the AUV will typically start outside this set.

To handle this, the docking task is split into two phases. Firstly, the vehicle is steered from its initial state toward the DS using dead reckoning (AHRS, DVL) aided by absolute position fixes (e.g., GPS or USBL relayed by a surface vehicle) and a standard path-following law. The AUV first converges to a homing region in the neighbourhood of the DS while keeping a safe depth. If acoustic communication is still unavailable upon arrival, it performs a circular search around the approximate DS location at fixed altitude close to the seabed, until the relative geometry between the vehicle and the docking station allows acoustic communication to be established. The phase transition requires several consecutive valid acoustic packets, preventing premature switching under intermittent packet loss.

\begin{figure}[b]
	\centering
	\includegraphics[width = 0.9\linewidth]{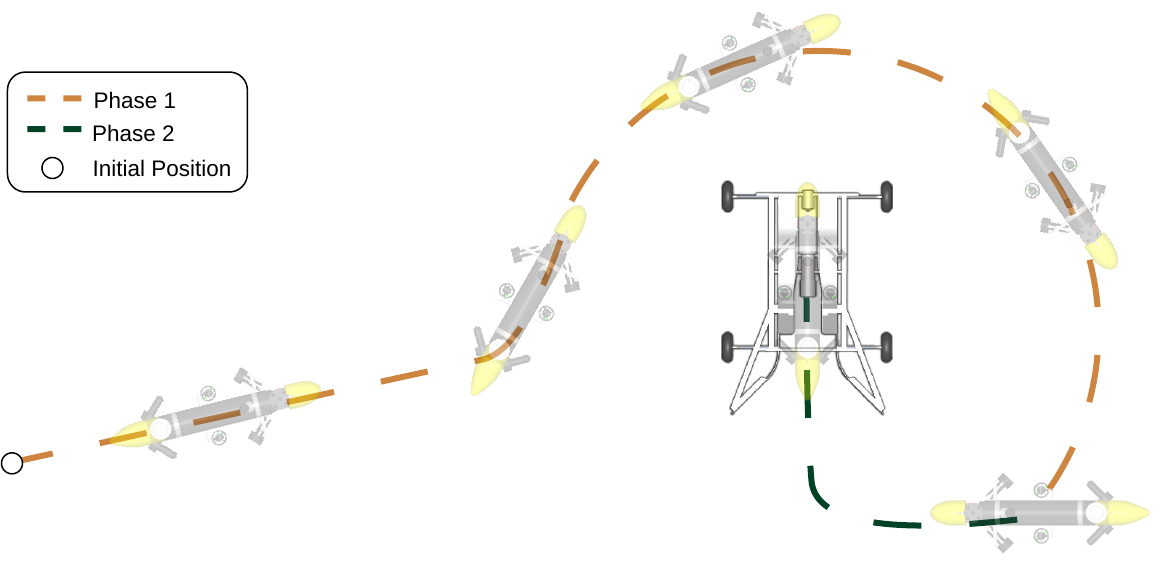}
	\caption{Planar overview of the proposed two-phase docking maneuver.}\label{fig:docking:state_machine}
\end{figure}

Once acoustic communications are reliably established, navigation becomes purely relative to the DS. The AUV estimates its pose in the DS frame using USBL measurements and a trajectory-tracking controller drives the vehicle along a terminal docking path. A diagram of the full manoeuvre is represented in Figure~\ref{fig:docking:state_machine}. The next subsection describes how this terminal trajectory is generated.

\subsection{Docking trajectory generation}


During the second phase, the final approach to the DS is executed along a terminal trajectory expressed in the $\mathcal{D}$-frame. This trajectory is generated once online, as soon as acoustic communications are established and an estimate of the relative pose is available. 
It is not continuously replanned, nor is it intended to optimise distance travelled. Instead, its purpose is to provide a safe, smooth, and feasible reference that the controller can reliably track during the final docking manoeuvre.

\begin{figure}[t]
	\centering
	\includegraphics[width = 0.9\linewidth]{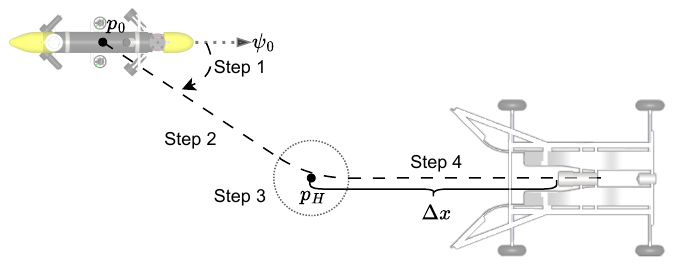}
	\caption{Diagram detailing the steps that compose the trajectory for the final docking manoeuvre.}\label{fig:docking:trajectory}
\end{figure}

The trajectory, represented in Figure~\ref{fig:docking:trajectory}, is constructed in four successive steps:
\begin{itemize}
    \item \textbf{Step 1}: the AUV first rotates in place until it faces the homing point $\bm{p}_H$ represented in Figure~\ref{fig:docking:trajectory};
    \item \textbf{Step 2}: it then moves forward in a straight line towards $\bm{p}_H$;
    \item \textbf{Step 3}: as the vehicle approaches $\bm{p}_H$, it gradually adjusts its heading so that it arrives aligned with the $x$-axis of the $\mathcal{D}$-frame;
    \item \textbf{Step 4}: finally, the AUV advances in a straight line along this axis and enters the DS;
\end{itemize}

In parallel, the $z$-coordinate converges to zero as fast as possible, independent from the horizontal motion. Pitch and roll references are held at zero and effectively unused, since the Medusa platform lacks actuation in these degrees of freedom. The Euler-angle references are converted into a desired rotation matrix so the resulting trajectory is defined in $SE(3)$ by the pair $(R_d, \bm{p}_d)$.
All components of the reference are generated using jerk limited curves, ensuring continuity in position, velocity, and acceleration. 

\section{Dual-USBL navigation system}\label{section:filter}
This section presents the navigation system for the final docking approach, using two filters: a complementary filter on $SO(3)$ for orientation and an Extended Kalman Filter for position. The next subsection introduces the orientation filter.

\subsection{Orientation filter}

The goal of the orientation filter is to estimate the rotation of the vehicle with respect to the docking station, represented by the matrix $_{\mathcal{B}}^{\mathcal{D}}R$. The adopted design follows the Explicit Complementary Filter of~\cite{mahony2008nonlinear}. For compactness, $R$ and $\hat{R}$ will denote the true and estimated rotations, respectively.

\begin{figure}[t]
	\centering
	\includegraphics[width = 0.8\linewidth]{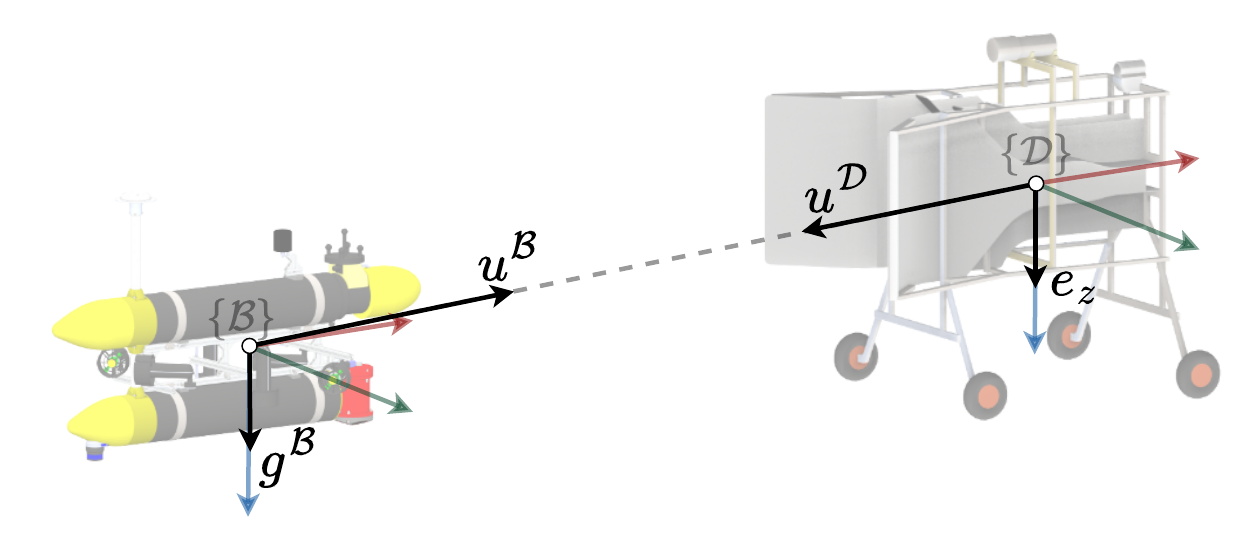}
	\caption{Geometry associated with the measurements used to drive the orientation filter.}\label{fig:filter:orientation}
\end{figure}

The seabed at the DS site is assumed to be horizontal, so that the $z$-axis of the $\mathcal{D}$ frame is aligned with the gravity direction. Two reference directions are therefore used: (i) the line-of-sight vector from the DS to the AUV, $\bm{u}^{\mathcal{D}}$, and (ii) the $\mathcal{D}$-frame $z$-axis, $\bm e_z$. Their representations in the body frame follow from the rotation $R$ as
\begin{equation}
    \bm{u}^{\mathcal{B}} = R^\top(-\bm{u}^{\mathcal{D}}), \qquad
    \bm g^{\mathcal{B}} = R^\top \bm e_z,
\end{equation}
where $\bm g^{\mathcal{B}}$ denotes the gravity direction measured by onboard sensors. The geometry associated with these measurements is represented in Figure~\ref{fig:filter:orientation}.

The filter dynamics are set as 
\begin{align}
    \dot{\hat{R}} &= \hat{R} \bigl(\varOmega - \hat{b} + k_P e_R \bigr)_\times, \qquad \hat{R}(0) = \hat{R}_0, \label{eq:mahony1} \\
    \dot{\hat{b}} &= -k_I e_R, \qquad \hat{b}(0) = \hat{b}_0, \label{eq:mahony2}
\end{align}
with correction term
\[
    e_R = k_1 \bigl( \bm{u}^{\mathcal{B}} \times \hat{R}^\top \bm{u}^{\mathcal{D}} \bigr)
        + k_2 \bigl( \bm g^{\mathcal{B}} \times \hat{R}^\top \bm e_z \bigr),
\]
where $\varOmega$ is the angular velocity of $\mathcal{B}$ with respect to $\mathcal{I}$, expressed in $\mathcal{B}$; $b$ is the gyroscope bias and $\hat{b}$ its estimate; and $k_P, k_I, k_1,$ and $k_2$ are positive gains.

The filter exhibits almost global asymptotic stability. A    detailed proof is provided in~\cite{mahony2008nonlinear}.

\subsection{Position filter}
The position filter estimates the position of the origin of $\mathcal{B}$ with respect to $\mathcal{D}$, expressed in $\mathcal{D}$, denoted $\bm p$. A complementary-style Extended Kalman Filter (EKF) is adopted. Although the gains are obtained through the standard EKF formulation using specified noise covariances, the resulting structure preserves the core principle of complementary filtering: DVL velocity measurements dominate at higher frequencies, while the USBL observations provide corrections at a lower rate.
\begin{figure}[b]
	\centering
	\includegraphics[width = 0.8\linewidth]{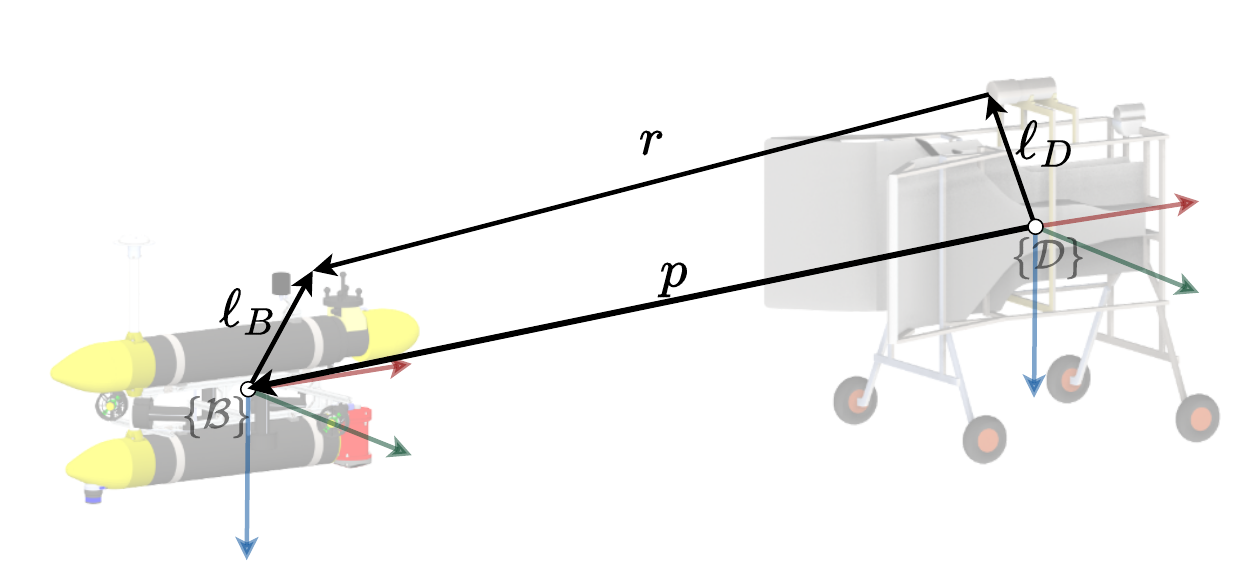}
	\caption{Geometry associated with the measurements used to drive the position filter.}\label{fig:filter:position}
\end{figure}

Let the state be $\bm{p} \in \mathbb{R}^3$, and the associated covariance $P \in \mathbb{R}^{3\times3}$. 
The discrete-time prediction equations are
\begin{subequations}
\begin{align}
    \hat{\bm{p}}_{k|k-1} &= \hat{\bm{p}}_{k-1|k-1} + \Delta t \, R \, \bm{V}_{k-1}, \label{eq:ekf_predict_x}  \\
    P_{k|k-1} &= P_{k-1|k-1} + \Delta t \, Q, \label{eq:ekf_predict_P}
\end{align}
\end{subequations}
where $R \in SO(3)$ is the vehicle orientation provided by the orientation filter, $\bm{V}
 $ is the velocity measurement expressed in $\mathcal{B}$, $\Delta t$ is the sampling period and $Q$ is derived from the measurement uncertainty associated with the DVL sensor, given by
\begin{equation}
    Q = \begin{bmatrix}
        \sigma_u^2 & 0 & 0 \\
        0 & \sigma_v^2 & 0 \\
        0 & 0 & \sigma_w^2  \\
    \end{bmatrix},
\end{equation}
where $\sigma_\star$ are the uncertainties associated with the measurement of each component of the body velocity.

The USBL sensors are not aligned with the vehicle or DS reference frames, but rather mounted at known offsets $\bm \ell_\mathcal{B}$ and $\bm \ell_\mathcal{D}$, respectively, as shown in Figure~\ref{fig:filter:position}. For the DS-mounted USBL, the measured geometry is 
\begin{equation}
    \bm{r}^\mathcal{D} = \bm{p} + R \: \bm{\ell}_B - \bm{\ell}_D.
\end{equation}
As a consequence, the measurement, which consists of the range and direction unit vector, is modelled as
\begin{align}
    h_D(\bm p) = \begin{bmatrix}  \|\bm{r}^\mathcal{D}\| \\[1pt] 
        \dfrac{\bm{r}^\mathcal{D}}{\|\bm{r}^\mathcal{D}\|}   \end{bmatrix}
    \, ,
\end{align}
with Jacobian
\begin{equation}
    H_{D} = 
    \dfrac{1}{\|\bm{r}^\mathcal{D}\|}
    \begin{bmatrix}
    (\bm{r}^\mathcal{D})^\top \\[1pt] 
    \mathbb{P}_{\bm u^\mathcal{D}}
    \end{bmatrix}.
\end{equation}

Similarly, for the vehicle-mounted USBL, the measured geometry is
\begin{equation}
    \bm r^\mathcal{B} = R^\top(\bm{\ell}_{D} - \bm{p}) - \bm{\ell}_{B} \,,
\end{equation}
and the corresponding measurement is modelled as 
\begin{align}
    h_B(\bm p) = \begin{bmatrix}  \|\bm{r}^\mathcal{B}\| \\[1pt] 
        \dfrac{\bm{r}^\mathcal{B}}{\|\bm{r}^\mathcal{B}\|}   \end{bmatrix}
    \, ,
\end{align}
with Jacobian
\begin{equation}
    H_\mathcal{B} = 
    \dfrac{1}{\|\bm{r}^\mathcal{B}\|}
    \begin{bmatrix}
    - (\bm{r}^\mathcal{B})^\top \; R^\top \\
    - \mathbb{P}_{\bm u^\mathcal{B}} \; R^\top
    \end{bmatrix}.
\end{equation}
The measurements from both USBLs are stacked into a single measurement vector with corresponding Jacobian
\begin{equation}
    h(\bm p) = \begin{bmatrix} h_{D} \\ h_{B} \end{bmatrix}, 
    \qquad 
    H = \begin{bmatrix} H_{D} & 0 \\ 0 & H_{B} \end{bmatrix}.
\end{equation}
The update equations are
\begin{subequations}
\begin{align}
    K_k &= P_{k|k-1} H_k^\top {(H_k P_{k|k-1} H_k^\top + \Sigma)}^{-1}, \label{eq:docking:ekf_k}  \\
    \hat{\bm{p}}_{k|k} &= \hat{\bm{p}}_{k|k-1} + K_k \left(\bm{z}_k - h(\hat{\bm{p}}_{k|k-1})\right), \label{eq:docking:ekf_update_state} \\
    P_{k|k} &= (I - K_k H_k) P_{k|k-1} {(I - K_k H_k)}^\top + K_k R K_k^\top,\label{eq:docking:ekf_update_cov}
\end{align}
\end{subequations}
where $\bm z_k$ is the sensor measurement correspondent to $h(\bm p)$ and $\Sigma$ is the measurement noise covariance given by
\begin{equation}
    \Sigma =
    \begin{bmatrix}
        \Sigma_\mathcal{D} & \bm{0} \\
        \bm{0} & \Sigma_\mathcal{B}
    \end{bmatrix},\qquad
    \Sigma_\star =
    \begin{bmatrix}
        \sigma_\rho^2 & \bm{0} \\
        \bm{0} &
        \sigma_{\beta}^2 \!\left(I - \bm{u}_\star \bm{u}_\star^\top\right)
    \end{bmatrix}.
\end{equation}
where $\sigma_\rho$ and $\sigma_\beta$ are the uncertainties associated with the measured range and angles respectively.

\subsection{Robust fusion of delayed acoustic measurements}



Outlier rejection is performed through innovation gating. In the position filter, each USBL update is accepted only if its Normalized Innovation Squared (NIS), computed from the innovation and its covariance, lies below a chosen $\chi^2$ threshold. For the attitude filter, the acoustic bearing residual is projected onto the tangent plane of the predicted unit vector on $S^2$, yielding a two-dimensional innovation to which an analogous $\chi^2$ gate is applied. Measurements outside the gate are discarded.

Delayed USBL measurements are handled with a fixed-lag buffering strategy. The filter stores recent states, covariances, and inertial measurements. When an out-of-sequence USBL measurement is received, the estimate is rolled back to the measurement timestamp, updated, and re-propagated to the current time using the buffered measurements. This ensures that each acoustic update is fused at the time to which it physically corresponds.

  \section{Trajectory-Tracking Controller}\label{section:control}
This section introduces the controller responsible for tracking the trajectory defined for the final docking manoeuvre.
A geometric \emph{inverse-dynamics} PID controller is proposed, formulated in $SE(3)$, inspired by the approach presented in~\cite{goodarzi2013geometric}.

Let $p\in\mathbb{R}^3$, $R\in SO(3)$, $v,\omega\in\mathbb{R}^3$,  $\bm{\eta}=(p,R)$, $\bm{\nu}=\begin{bmatrix}v\\ \omega\end{bmatrix}\in\mathbb{R}^6$, and $S(\cdot)$ the skew map so that $S(a)b=a\times b$ and $S^{-1}(\cdot)$ the inverse map. Consider the kinematics and dynamics equations in the form 
\begin{align}
  (\dot p\, , \dot R) &= (R\,v \, ,  R \, S(\omega)),\label{eq:kin}\\
  M\,\bm{\dot\nu} + C(\nu)\nu + D(\nu)\nu + g(\bm{\eta}) &= \bm{\tau}.\label{eq:dyn}
\end{align}
Consider also arbitrarily smooth inertial references $p_d(t)\in\mathbb{R}^3$ and $R_d(t)\in SO(3)$ be $C^2$ and bounded. Define the correspondent desired signals in the body frame as
\begin{align}
v_d &= R_d^\top \dot p_d, 
& \dot v_d &= R_d^\top \ddot p_d - S(\omega_d)\,v_d, \label{eq:vd-vdotd}\\
S(\omega_d) &= R_d^\top \dot R_d,
& S(\dot\omega_d) &= R_d^\top \ddot R_d - S{(\omega_d)}^2. \label{eq:wdd}
\end{align}

Let $e_p$ and $e_v$ be position and velocity tracking errors and $e_R$, $e_\omega$ the attitude and angular velocity tracking errors, expressed in the body frame, given by
\begin{align}
e_p &= R^\top(p - p_d) \in \mathbb{R}^3, \label{eq:eb}\\
e_v &= v - R^\top R_d\,v_d \in \mathbb{R}^3, \label{eq:ev}\\
e_R &= \tfrac{1}{2}\,S^{-1}\!\big(R_d^\top R - R^\top R_d\big)\in\mathbb{R}^3, \label{eq:eR}\\
e_\omega &= \omega - R^\top R_d\,\omega_d \in \mathbb{R}^3. \label{eq:ew}
\end{align}
The following kinematic identities are derived from~\eqref{eq:kin}-\eqref{eq:wdd}:
\begin{align}
\frac{d}{dt}\big(R^\top R_d\,v_d\big) &= R^\top \ddot p_d - S(\omega)\,R^\top R_d\,v_d, \label{eq:transport_vd} \\
\frac{d}{dt}\big(R^\top R_d\,\omega_d\big) &= R^\top R_d\,\dot\omega_d - S(\omega)\,R^\top R_d\,\omega_d. \label{eq:transport_wd}
\end{align}

The commanded accelerations $\dot v^\star$ and $\dot\omega^\star$ are obtained
from PID-like and PD-like laws with acceleration feedforward. Integral action is
used only in the position controller to reject slowly varying disturbances, such as ocean
currents, with
\begin{align}
  \dot z_p &= e_v + c_{ip} e_p,
\end{align}
where $c_{ip}>0$.
The desired error dynamics are set as
\begin{align}
\dot e_v + K_{dv} e_v + K_{pp} e_p + K_{ip} z_p &= 0, \label{eq:evdyn}\\
\dot e_\omega + K_{d\omega} e_\omega + K_{pR} e_R &= 0, \label{eq:ewdyn}
\end{align}
where $K_{pp},K_{dv},K_{ip}\in\mathbb{R}^{3\times 3}$ and $K_{pR},K_{d\omega}\in\mathbb{R}^{3\times 3}$ are diagonal, positive-definite gains. 

From~\eqref{eq:ev} and~\eqref{eq:transport_vd}:
\begin{equation}
  \dot e_v = \dot v - \big(R^\top \ddot p_d - S(\omega)R^\top R_d v_d\big).
\end{equation}
Enforcing~\eqref{eq:evdyn} gives the translational virtual acceleration:
\begin{equation}
  \dot v^\star = R^\top \ddot p_d - S(\omega)R^\top R_d v_d
                - K_{dv} e_v - K_{pp} e_p - K_{ip} z_p.
  \label{eq:vdotstar}
\end{equation}
Similarly, from~\eqref{eq:ew} and~\eqref{eq:transport_wd}:
\begin{equation}
  \dot e_\omega = \dot\omega - \big(R^\top R_d \dot\omega_d - S(\omega)R^\top R_d \omega_d\big),
\end{equation}
and enforcing~\eqref{eq:ewdyn} yields
\begin{equation}
  \dot\omega^\star = R^\top R_d \dot\omega_d - S(\omega)R^\top R_d \omega_d
                    - K_{d\omega} e_\omega - K_{pR} e_R \,.
  \label{eq:omegadotstar}
\end{equation}

The final step to obtain the control law is to use the dynamics model to compute the forces and torques that shall be commanded to the vehicle. 
This strategy is suitable given that the Medusa vehicle model has been previously identified and its accuracy experimentally verified.

Letting $\dot\nu^\star=\begin{bmatrix} \dot v^\star\ \\ \dot\omega^\star \end{bmatrix}$, the control is given by
\begin{equation}
  \bm{\tau} \;=\; M\,\dot{\bm{\nu}^\star} \;+\; C(\bm{\nu})\bm{\nu} \;+\; D(\bm{\nu})\bm{\nu} \;+\; g(\bm{\eta}). \label{eq:computed_torque}
\end{equation}


Under suitable gain conditions, the controller achieves global asymptotic stabilization of the position error and almost-global asymptotic stabilization
of the attitude error, except for the measure-zero set of unstable attitude critical points on $\mathrm{SO}(3)$. The proof follows standard Lyapunov arguments for geometric tracking controllers on $\mathrm{SE}(3)$, in line with~\cite{goodarzi2013geometric}, and is omitted here due to space constraints.

  \section{Results}\label{section:results}
This section presents the key results obtained from software implementation and testing of the proposed algorithms in realistic simulations and real-world experiments.

\subsection{Simulation results}
\begin{figure}[b]
    \centering
    \includegraphics[width=\linewidth]{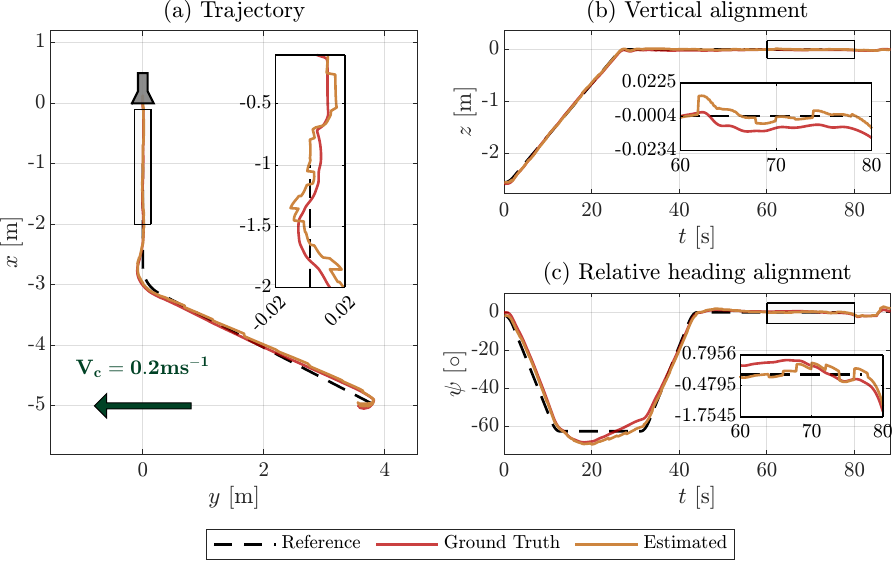}
    \caption{Overview of the controller's performance using the estimated state, under the effect of an ocean current.}\label{fig:results:controller_filter_overview_current}
\end{figure}

The performance of the proposed filter and controller during the final docking manoeuvre was evaluated under a uniform current of \SI{0.2}{\metre\per\second} with a heading of $-90^{\circ}$ relative to the DS. 
As shown in Figure~\ref{fig:results:controller_filter_overview_current}, the estimated state closely follows the real one and the planned trajectory is successfully tracked, leading to a successful manoeuvre.

\subsection{Real-world trials}
\begin{figure}[t]
    \centering
    \includegraphics[width=0.8\linewidth]{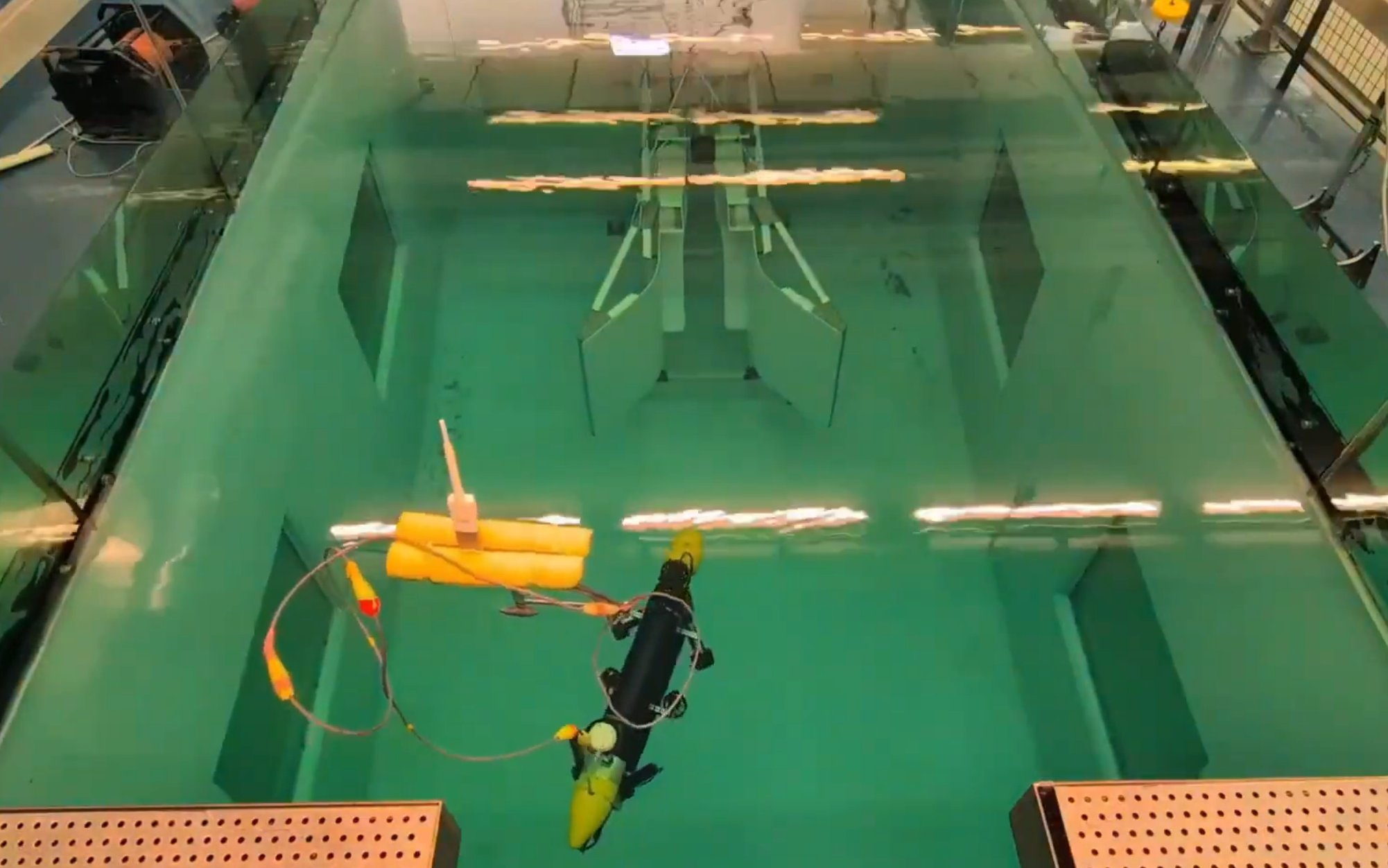}
    \caption{The testing site at IST - Taguspark. The real Medusa vehicle and the DS can also be seen.}\label{fig:results:tagus}
\end{figure}

The ultimate goal of this work is to validate the proposed system on the real Medusa vehicle. The experiments were conducted in a $5$by$3$\SI{}{\metre} water tank with a depth of \SI{4}{\metre}, located at the IST Taguspark campus, depicted in Figure~\ref{fig:results:tagus}.

Testing in the water tank poses major challenges for an acoustic-based system. 
The confined environment causes severe multipath due to the sound reflecting on the walls and water surface. This means that reliable communication is only possible in a small region where the vehicle's USBL is aligned with the DS boresight. 
As a result, each run had to start from a favourable position to ensure initial acoustic communication and prevent filter divergence.

\begin{figure}[t]
    \centering
    \includegraphics[width=\linewidth]{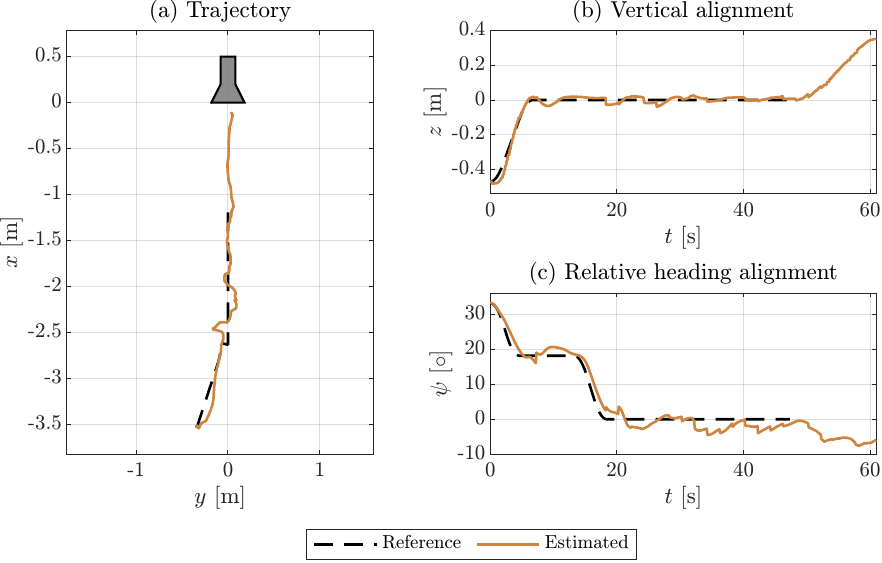}
    \caption{Overview of the full docking system's performance on the real Medusa vehicle.}\label{fig:results:controller_real_filter_overview_current}
\end{figure}

In each experiment, the vehicle was initialized as far from the DS as possible and with an intentional misalignment.
Results from one such experiment are shown in Figure~\ref{fig:results:controller_real_filter_overview_current}.
It can be confirmed that the reference trajectory is accurately generated and tracked, confirming the success of the docking manoeuvre. 
To improve acoustic coverage, the origin of $\mathcal{D}$ was slightly shifted upward, promoting better alignment of the two USBL sensors. This explains the increase in the $z$ coordinate as the vehicle enters the DS.
The experiment was repeated five consecutive times, all of them successful.

  \section{Conclusions}\label{section:conclusion}

This paper presented a fully acoustic solution to the underwater docking problem, offering clear advantages over the camera-based approaches that dominate the literature. The proposed navigation architecture combines a nonlinear complementary filter for attitude with an Extended Kalman Filter for relative position, enabling accurate localization with respect to the docking station during the terminal phase. A geometric trajectory-tracking controller ensures that the vehicle is safely guided into the dock.

The complete architecture was validated in high-fidelity simulations and experimentally demonstrated on the Medusa platform, confirming the feasibility and robustness of the proposed docking solution.


Future work will focus on strengthening the stability properties of the tracking controller and improving robustness to model uncertainty.

  \bibliography{ifacconf}             








  \end{document}